# High frequency single-mode resonators for EPR spectroscopy enabling rotations of the sample about two orthogonal axes

G. Annino[1, @], M. Cassettari[1], M. Fittipaldi[2], and M. Martinelli[1]

[1] *Istituto per i Processi Chimico-Fisici, Area della Ricerca CNR, via G. Moruzzi 1, 56124 Pisa (Italy)*
[2] *Departement Natuurkunde, Universiteit Antwerpen, Universiteitsplein 1, B-2610 Antwerpen (Belgium)*

(April 22, 2005)

**Abstract**

A novel single-mode resonant structure which enables the rotation of the sample about two orthogonal axes is investigated in view of electron paramagnetic resonance applications. The proposed solution is based on cylindrical nonradiative resonators laterally loaded by the sample holder. The resulting structure can still operate in nonradiative regime, although no longer rotationally invariant. These theoretical predictions, based on symmetry considerations, are confirmed by means of a finite element numerical modelling. Theoretical and computational results are then substantiated by experimental investigations at millimeter wavelengths. As a result, a single-mode resonator which enables all the relevant rotations of the sample is demonstrated at millimeter wavelengths for the first time. In this resonator the intensity of the microwave field on the sample and its orientation with respect to the static magnetic field can be kept constant during the rotations. Therefore, a complete characterization of anisotropic systems is possible at the highest sensitivity, without the need of split-coil magnets. Possible applications at very high frequencies are discussed.

---

*e-mail address: geannino@ipcf.cnr.it*

# 1. Introduction

Among the different components which constitute a typical Electron Paramagnetic Resonance (EPR) spectrometer, the resonator plays a fundamental role. Indeed, it increases the electromagnetic energy density and decouples the electromagnetic field interacting with the sample from that of the propagation circuit. This is especially true at millimeter wavelengths, where the available power and the level of standing waves are often too far from the desired values. Possible limitations in the use of this key component are imposed by the constraints of a specific measurement technique or by the high working frequency, often by the combination of both these factors.

A demanding application of EPR spectroscopy is the analysis of single crystals, which would require in general a complete rotation of the sample and the use of open cavities. This latter possibility allows an effective static magnetic field modulation and the radiofrequency excitation in case of electron-nuclear double resonance. Open cavities are also necessary for optically-detected magnetic resonance spectroscopy. The realization of a resonator allowing these capabilities can be complicated by the high working frequency. For instance, in case of single crystals of proteins, in which the maximum size of the sample can be very tiny, the use of a millimeter wave spectrometer is dictated by the necessary absolute sensitivity.

At millimeter wavelengths the most common resonant devices are the Fabry-Perot cavities [1-4]. The structure of these overmoded resonators is intrinsically open, so that they represent a good candidate for the implementation of the rotation of the sample [5] and of additional electromagnetic irradiations [6]. The quality factor $Q$ of a Fabry-Perot can be extremely high. However, at millimeter wavelengths it is typically limited to few thousands when the overall size of the resonator is of the order of few wavelengths. The absolute sensitivity of an EPR spectrometer is proportional to the intensity of the microwave magnetic field on the sample, and thus to the ratio between the merit factor $Q$ and the active volume of the resonator. The pulse width and the time resolution achievable in pulsed spectroscopy are also limited by the amplitude of the microwave field on the sample. In this respect, single-mode resonators can guarantee higher performances, as well as a lower level of field distortion and diffraction losses due to the sample. These reasons have motivated the efforts to realize millimeter wave standard $TE_{011}$ cylindrical cavities. The main difficulties to overcome in this case are the excitation configuration and the close structure. The solution commonly adopted for the excitation of this kind of resonator is the coupling hole. At 100 GHz the typical diameter of this aperture is about 0.8 mm; the corresponding thickness of the cavity wall around the hole is less than 0.05 mm. A standard TE$_{011}$ cavity working at this frequency can be effectively opened to other electromagnetic irradiations by cutting several slits of 0.35 mm width perpendicularly to its axis, as shown in Ref. [7]. More recently the same approach has been successfully pursued for the realization of a 275 GHz cavity [8, 9]. In such resonators, the implementation of sample rotations about two orthogonal axes seems quite problematic. The adopted solution is indeed limited to rotations about a single axis [10], or based on split-coil magnets. In the latter case, the relative orientation between the static magnetic field $\vec{B}_0$ and the g-tensor can be varied rotating the sample about the axis of the cavity and $\vec{B}_0$ about another orthogonal axis [11]. The intrinsic drawback of this arrangement is the dependence on the angles of rotation of the transition probability between Zeeman levels. This probability varies with the relative orientation between the static magnetic field and the microwave field, and vanishes when these fields are aligned. The practical drawback is the necessity of a split-coil magnet, still not available at very high fields.

The development of novel millimeter wave single-mode resonators characterized by inherent ease of realization and open structure requires a different strategy. A recent proposal in this direction is given by the partially open dielectric resonators working on the so-called NonRadiative (NR) configuration [12, 13]. This solution allowed the achievement of the state of the art of the power-to-field conversion efficiency at room temperature. The same NR principle was extended to metallic resonators. By this way a TE$_{011}$ cavity combining the open structure typical of a Fabry-Perot with the conversion efficiency of a single-mode resonator was obtained [14]. The concept of NR resonator was then generalized in order to include axially open configurations, in which the sample holder acts itself as single-mode resonant element [15]. The unifying aspect of these devices is their working principle, which ensures a partially open structures and a simple excitation configuration. As a consequence, all the mentioned NR resonators are potential candidates for the realization of a millimeter wave single-mode resonator which enables rotations of the sample about orthogonal axes. The basic aim of this work is the theoretical and experimental investigation of this possibility. The recognition of possible resonant configurations allowing the requested rotations will be the starting point of this investigation. The NR principle will be generalized to structures without rotational invariance on the basis of different symmetry properties. The theoretical predictions will be first supported by a computational analysis and then substantiated by experimental characterizations made around 70 GHz. The obtained results represent to our knowledge the first demonstration of single-mode resonators enabling a complete orientation study of the sample at millimeter wavelengths, without the need of a split-coil magnet or of a controlled remounting of the sample.

This paper is structured as follows. In Sect. 2, the physical background underlying the NR resonators is introduced. A possible variant allowing sample rotations is presented and its design criteria are discussed. In Sect. 3, the expected field distribution of practical configurations is shown as calculated by means of a finite element method. The experimental characterization of these configurations is reported in Sect. 4. Sect. 5 is dedicated to the discussion of the obtained results. The perspectives which they open for applications at very high frequencies are outlined as well. Finally, the conclusions are summarized in Sect. 6.

## 2. Theoretical aspects

In a typical NR device, a central region is surrounded by a partially open structure given by two ideally infinite parallel conducting plates kept at distance *l*, as shown in Fig. 1 for the case of a dielectric disc. When the free space wavelength $\lambda_0$ of the employed radiation fulfils the NR condition $l < \frac{\lambda_0}{2}$, only the cutoff-less TEM mode of the parallel-plates structure can propagate outside the central region. The propagation of the TE and TM modes is indeed prevented by their cutoff condition, as discussed in Ref. [16] in terms of plane waves. A normal mode of the central region is thus unaffected by irradiation losses when its characteristic frequency fulfils the NR condition and its projection on the TEM mode is negligible [14]. The central region of the NR device can therefore exhibit confined resonance modes also in presence of a partially open structure. In specific cases the existence of such nonradiating modes can be argued on the basis of symmetry considerations. For instance, in a structure with circular symmetry about a particular axis, the modes with azimuthal invariance about this axis are necessarily transverse-electric TE or transverse-magnetic TM [14]. The former family of modes, hereafter referred to as $TE_0$, is characterized by an azimuthal electric field and by a combination of axial and radial magnetic fields. The $TE_0$ modes do not share any field component with the TEM mode, which is characterized only by an axial electric field and by an azimuthal magnetic field. Accordingly, at least a nonradiating family of modes is expected in rotationally invariant configurations.

The NR resonators can be excited by means of a simple configuration exploiting their NR character [12]. Indeed, a radiation incident on the NR structure is totally reflected back when its polarization is parallel to the conducting plates. Only an evanescent field extends inside the plates. This field can transfer energy to the resonator when a sufficient overlap with the field of the resonance mode exists. The level of coupling can be adjusted varying for instance the distance between the incoming radiation and the central region of the resonator.

The cylindrical dielectric configuration shown in Fig. 1 can be considered the ancestor of any NR resonator. It represents in fact a natural development of the NR waveguide first proposed [17, 18]. The radiation can be stored in the central region of this structure since the dielectric material increases the optical distance between the metallic plates [12, 14]. The $TE_{011}$ mode of this resonator shows the typical magnetic dipole field distribution, in which the maximum magnetic field $\vec{B}_\mu$ lies in the center of the resonator and is directed along its axis, as shown in Fig. 1. The relative position of the excitation waveguide with respect to the resonator is shown in Fig. 1 as well, together with a possible orientation of the static magnetic field $\vec{B}_0$.

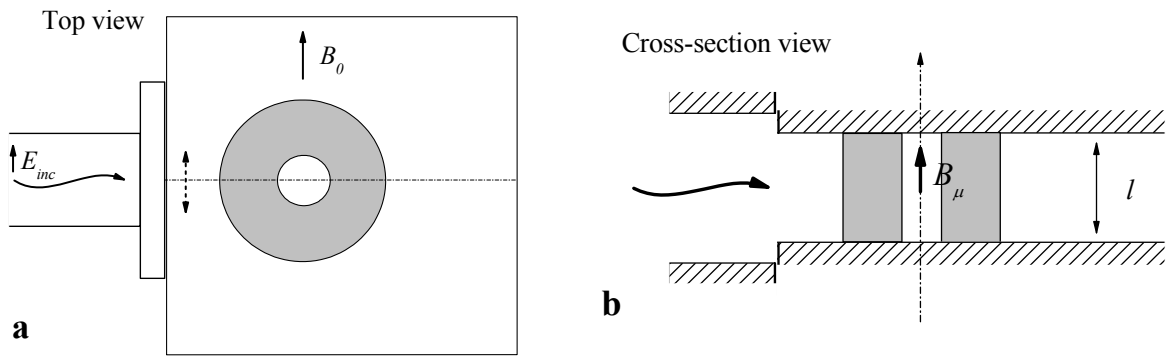

Fig. 1. Top view (**a**) and cross-section view (**b**) of a NR dielectric resonator with thickness *l*. The central hole can contain the sample. The linear polarization of the incoming radiation is shown in **a**. The maximum magnetic field $\vec{B}_\mu$ of the $TE_{011}$ mode and a possible orientation of the static magnetic field $\vec{B}_0$ are shown in **b**. The coupling to the incoming radiation can be changed by moving the resonator as indicated by the dashed line in **a**.

The insertion of the sample in resonators working on the $TE_{011}$ mode is typically realized through a cylindrical sample holder placed along the axis of the resonator. This solution is quite convenient since ensures a limited microwave field distortion and an easy sample rotation about the axis of the cavity. However, the geometry of the resonator prevents in general the rotation about a second orthogonal axis. To overcome this difficulty the insertion of the sample holder can be done laterally, namely with the axis of the capillary orthogonal to that of the resonator. This solution is shown in Fig. 2. The sample holder can be still rotated about its own axis (angle $\psi$ of Fig. 2b). In addition, the resonator as a whole can be rotated about its original axis (angle $\theta$ of Fig. 2a, considered with respect to $\vec{B}_0$), which is orthogonal to the axis of the sample holder and to the static magnetic field.

Fig. 2b shows a variant of the basic NR resonator of Fig. 1. In this case the conducting plates have been modified including two metallic plungers in correspondence of the dielectric region. The working principle and the basic characteristics of the resonator remain the same, whereas the plungers ensure the tuning of the resonance mode.

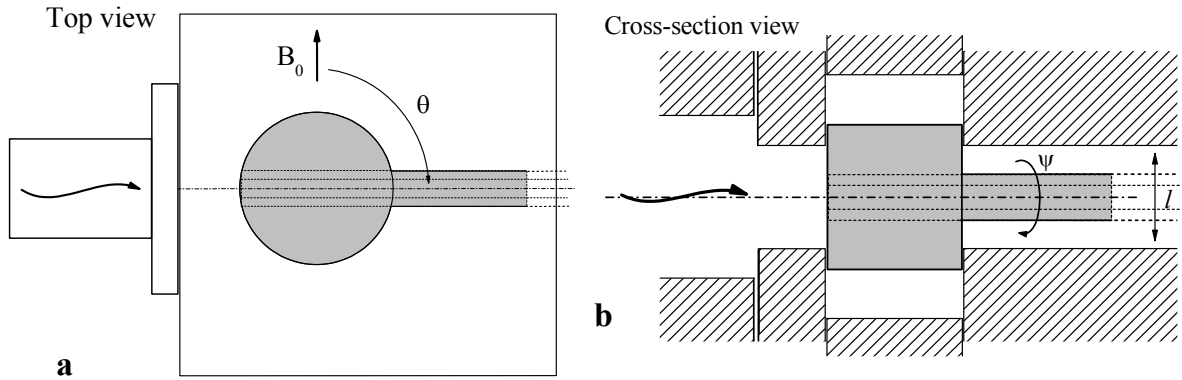

Fig. 2. Top view (**a**) and cross-section view (**b**) of a NR dielectric resonator laterally loaded by the sample holder. The angle $\theta$ indicates the orientation of the sample holder with respect to the direction of $B_0$, the angle $\psi$ the rotations about its axis. The planes of symmetry are indicated by dash-dotted lines.

By considering a reference frame with the z-axis along $\vec{B}_0$, it is possible to recognize in $\theta$ and $\psi$ the second and third Euler angles as introduced in Ref. [19], respectively. The first Euler angle corresponds in the present case to rotations about $\vec{B}_0$. The combination of rotations in $\theta$ in the interval $(0, \pi)$ and of rotations in $\psi$ in the interval $(0, 2\pi)$ covers all the significant orientations of the sample, being the rotations about $\vec{B}_0$ irrelevant for magnetic resonance purposes [20].

The configuration of Fig. 2 seems appropriate for a complete orientation analysis of the sample as well as for additional irradiations. The optical access to the sample can be obtained either through the sample holder or through a central axial hole in the dielectric disc. The fundamental question concerns the effect of the sample holder on the electromagnetic properties of the resonator. The new configuration loses indeed the rotational invariance, which was the crucial point in the prediction of a proper NR resonant behaviour [14]. Moreover, for practical millimeter wave applications the diameter of the sample holder cannot be much smaller than the size of the resonator. As a consequence it cannot be considered *a priori* a negligible perturbation of the field distribution.

Although the rotational invariance is no longer fulfilled, the NR behaviour of the structure of Fig. 2 can be still demonstrated on the basis of other symmetry properties. In particular, this structure is characterized by a specular symmetry with respect to the median plane of the resonator. The field distribution outside an arbitrary cylindrical region including the dielectric disc and the sample holder will be first considered. A complete basis of modes in this parallel-plates region is given by TEM, TE and TM modes, according to Ref. [16]. In cylindrical coordinates these modes can be labelled by an azimuthal modal index $n$ and by an axial index $m$, where $n$ indicates the number of azimuthal oscillations and $m$ the number of axial oscillations. In this representation the TEM mode corresponds to the mode $TM_{n=0,m=0}$. In general each mode is characterized by a cutoff condition related to its axial index $m$. When expressed in terms of the distance between the conducting plates, it results that $l_{cutoff,m} = m\frac{\lambda_0}{2}$. As a consequence, the modes with vanishing axial index are cutoff-less. The symmetry of the structure allows a separation of the different modes according to their parity with respect to the median plane. In particular, the electric and magnetic fields can be chosen with even or odd parity with respect to this plane. Depending on the specific symmetry of the mode, the median plane can be ideally replaced by a Perfect Magnetic Conductor (PMC), in case of even axial magnetic field and odd axial electric field, or by a Perfect Electric Conductor (PEC) in the opposite case [21]. Therefore, the modes propagating between the conducting plates outside the central region can be grouped in a class of modes compatible with a median PEC plane and a class of modes compatible with a PMC plane. The PEC class is constituted by the modes $TM_{0,0} \equiv TEM$, $TM_{n,0}$ and $\{TE_{n,m}, TM_{n,m}\}_{m\ even\neq 0}$, whereas the PMC class is given by the modes $\{TE_{n,m}, TM_{n,m}\}_{m\ odd}$. The PEC class contains modes without cutoff, whereas the lowest cutoff of the PMC class is given by the condition $l_{cutoff} = \frac{\lambda_0}{2}$.

Following the same arguments as above, the normal modes of the central region of the resonator can be separated in modes compatible with a PMC median plane and modes compatible with a PEC plane. A PMC mode of the central region cannot couple to a PEC mode of the external region and vice versa, since the opposite parity makes them orthogonal to each other [23]. As a consequence, it is possible to conclude that the PMC modes of the central region of a NR structure having a median plane of symmetry are necessarily confined whenever the wavelength $\lambda_0$ associated to their characteristic frequency fulfils the NR condition $l < \frac{\lambda_0}{2}$ [24]. The family of the NR structures includes therefore the vast class of configurations characterized by the planar symmetry alone.

These general considerations can be applied to the case of interest here. The starting point is the NR resonator without the sample holder. This structure is characterized by circular and planar symmetry. Its $TE_{011}$ mode belongs to the PMC class. The inclusion of the lateral hole and of the sample holder removes the circular symmetry, perturbing the original $TE_{011}$ field distribution and resonance frequency. The resulting mode is still a PMC mode, and it can be confined provided that its final frequency fulfils the NR condition. Accordingly, a suitable choice of the initial resonance frequency and of the distance between the conducting plates guarantees a proper resonance also in the configuration of Fig. 2.

Although not strictly necessary, further properties of the resonance mode are highly desirable in view of effective EPR applications. First of all the part of the resonator accommodating the sample should be characterized by a minimum of electric field and a maximum of axial magnetic field, in order to minimize the dielectric losses and to maximize the EPR sensitivity. The evanescent field around the dielectric region should be only weakly dependent on the azimuthal angle, in order to ensure a relatively constant coupling level for different $\theta$ angles. For these practical reasons, the final field distribution should still be as much as possible $TE_{011}$-like.

### 3. Numerical modelling

The electromagnetic field distribution, resonance frequency, and quality factor of complex geometries can be obtained with the aid of a finite element numerical method. The configurations investigated in this paper were modelled by means of the FEMLAB 3.1 software (COMSOL, Sweden). A basic parameter of this numerical approach is the amount of elementary regions defined in the volume of analysis. Better accuracies are in general obtained for a large number of these elementary regions. Since this number is limited by practical reasons, it is convenient to reduce the volume of analysis as much as possible. This can be done by exploiting the symmetry of the problem. In the present case the symmetry with respect to the median plane can be employed to reduce the modelling to the upper half of the resonator. The plane of symmetry can be replaced by a PMC, as discussed. An additional plane of symmetry includes the axis of the original resonator and that of the capillary, neglecting the presence of the excitation configuration. This plane can be replaced by a PEC wall or by a PMC wall, depending on the mode of interest. The final geometry of analysis is shown in Fig. 3 and represents a quarter of the actual resonator.

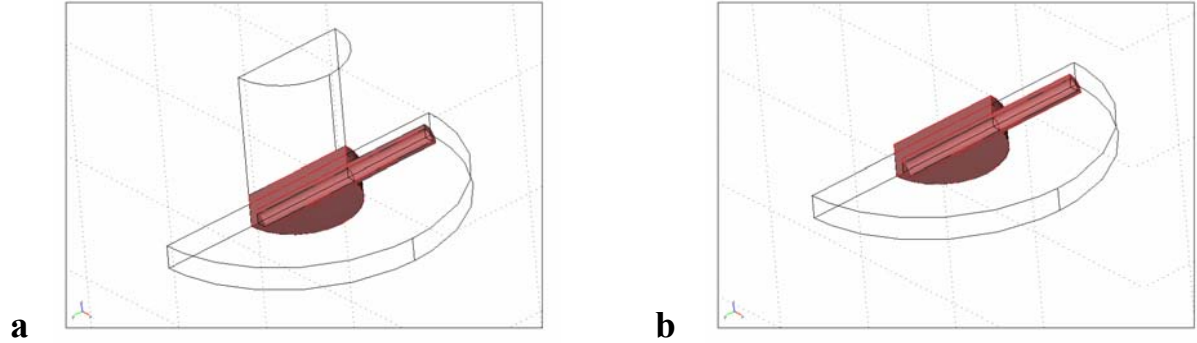

Fig. 3. Sections of the NR resonator investigated with the finite element method. **a** Configuration with the plungers 4 mm far from the plane surfaces of the dielectric disc. **b** Configuration with the plungers in contact with the plane surfaces of the disc. The dielectric regions are indicated by the shaded area.

Fig. 3a shows in particular a configuration in which the plungers are 4 mm far from the plane surfaces of the dielectric disc (considered of Teflon, $\varepsilon = 2.05$). Fig. 3b shows a configuration in which the plungers are in contact with the disc. In both cases the capillary (considered of Suprasil, $\varepsilon = 3.8$) crosses almost completely the disc, and extends well beyond its rim. The Teflon disc has a diameter of 3.7 mm and a thickness of 2.1 mm. The cylindrical holes in the conducting plates have the same diameter of the dielectric disc. The distance between the plates is 1.5 mm. In this case the expected resonance frequency of the unperturbed $TE_{011}$ mode, calculated assuming a homogeneous disc and no capillary, is 67.58 GHz as calculated with the FEMLAB software. The related half wavelength, $\frac{\lambda_{0,TE_{011}}}{2} = 2.22$ mm, is well beyond the distance between the conducting plates, as requested by the NR condition. Accordingly, the presence of the capillary should not push the final resonance frequency outside the allowed values. In the case here investigated the sample holder is a Suprasil tube with outer and inner diameters of 0.9 mm and 0.6 mm, respectively, which represents a typical choice for W-band measurements [11]. The frequency of the perturbed $TE_{011}$ mode of the resonator loaded by the sample holder is expected at 67.26 GHz. The calculated distributions of the norm of the electric field and of the axial magnetic field are shown in Fig. 4a and 4b, respectively.

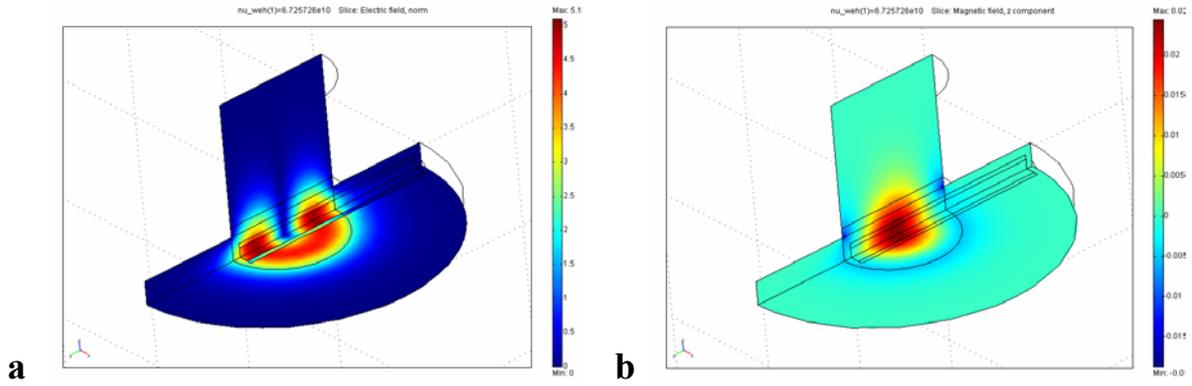

Fig. 4. Calculated norm of the electric field (**a**) and amplitude of the axial magnetic field (**b**) for the configuration with the plungers 4 mm far from the plane surfaces of the dielectric disc. The colours scale is shown on the right of the figures.

The electromagnetic energy is well confined in the central dielectric region, in agreement with the above theoretical analysis. The field distributions still have an approximate rotational invariance and are characterized by a node of electric field and an antinode of magnetic field in the center of the structure. From the point of view of EPR applications the resulting resonance can be properly indicated as a 'quasi-TE$_{011}$' mode. The numerical analysis in a large frequency interval centered on the quasi-TE$_{011}$ mode reveals no additional close resonances. It is worthwhile to note that the effect of the plungers in this configuration is negligible, due to the rapid axial decrease of the electromagnetic fields. Indeed, the replacement of the plungers with perfectly matched layers, which emulate the behaviour of the free-space propagation [25], does not affect resonance frequency and field distribution of the quasi-TE$_{011}$ mode.

A similar analysis was developed for the configuration in which the plungers are in contact with the disc. The behaviour of the resonator in this condition is of large practical importance, since the resonance frequency needs to be adjusted to compensate the thermal expansion and the detuning induced by the sample. In this configuration the resonance frequency of the initial TE$_{011}$ mode is expected at 74.47 GHz, about 10% above the value obtained in absence of the plungers. The new field distributions are reported in Fig. 5. In particular Fig. 5a shows the norm of the electric field and Fig. 5b the amplitude of the axial magnetic field. Both distributions are still well confined, as expected. Their behaviour along the azimuth of the disc remains approximately homogeneous, although less than in the previous case, and still characterized by a node of electric field and an antinode of magnetic field in the center of the resonator. The resulting mode can be therefore considered as quasi-TE$_{011}$. The expected spectrum around this resonance is however denser, and the closest resonance lies at about 1 GHz.

In both the investigated configurations the direction of the microwave magnetic field in the center of the resonator is basically axial. The angle between this magnetic field and the static one is thus independent of the rotation angles.

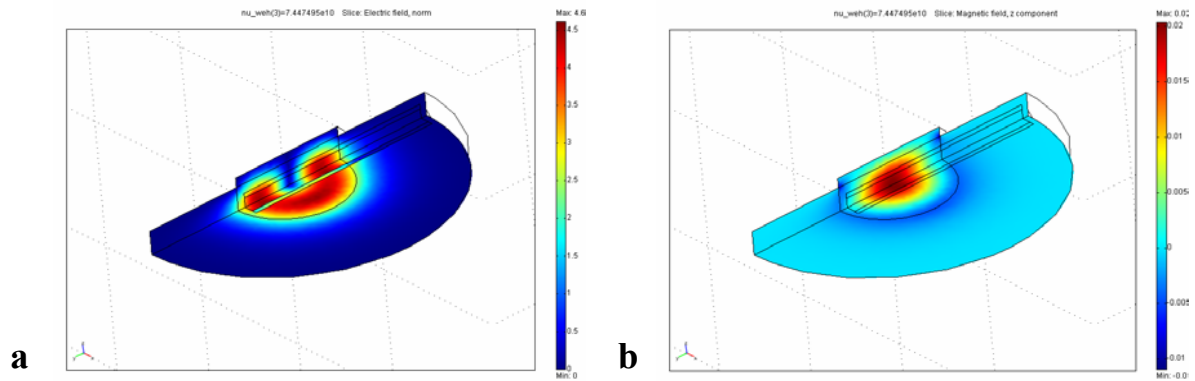

Fig. 5. Calculated norm of the electric field (**a**) and amplitude of the axial magnetic field (**b**) for the configuration with the plungers in contact with the plane surfaces of the dielectric disc. The colours scale is shown on the right of the figures.

### 4. Experimental evidences

The experimental confirmation of the previous theoretical and numerical predictions was based on Teflon discs with a diameter of $3.6 \pm 0.05$ mm and a thickness of $2.1 \pm 0.05$ mm. The diameter of the lateral hole, which crosses the disc from side to side along its center, was $0.9 \pm 0.05$ mm. This hole accommodated a Suprasil capillary with nominal outer and inner diameters of 0.84 mm and 0.6 mm, respectively. The conducting mirrors were realized with aluminium plates with a thickness of 6 mm, fixed at a distance of $1.5 \pm 0.05$ mm. The diameter of the axial holes was $3.7 \pm 0.05$ mm. The distance between the rim of these holes and that of the plates was $1.1 \pm 0.05$ mm. The diameter of the plungers, realized with aluminium, was $3.6 \pm 0.05$ mm. The position of the dielectric region along the holes in the metallic plates was kept stable by means of the capillary.

The resonators were characterized in the frequency interval from 55 to 92 GHz. Their excitation was obtained through a reflection configuration by using a WR-15 V-band rectangular waveguide. A detailed description of the employed setup and of the procedure used to normalize the resonance curves can be found in Ref. [12].

In the first series of measurements the plungers were not used, so that the structure of the resonator resembled the axially open configuration discussed in Ref. [15]. The identification of the resonance modes was first done on a homogeneous disc, in absence of lateral hole and sample holder. The resulting spectrum of resonance frequencies is given by the upper curve of Fig. 6. The $TE_{011}$ mode is easily identified, since the resonance appearing at 66.65 GHz corresponds quite well with the value of 67.58 GHz calculated for this mode, at least with respect to the distance between different modes. The residual discrepancy between the calculated frequency and the experimental one can be ascribed to the differences between the ideal configuration and that actually employed in the measurements. Several other modes were observed and identified with the aid of the numerical modelling. The classification of these secondary resonances was done on the basis of their symmetry with respect to the median plane, following the nomenclature proposed in Ref. [26]. In addition to the $TE_0$ modes, in which the indices are used in a standard way, the other modes are here labelled with $SX_n$ or $NX_n$. The first letter indicates the parity of the axial magnetic field with respect to the median plane. The label S refers to symmetric (even parity) modes and the label N to antisymmetric (odd parity) modes. The index $n$ corresponds to the already introduced azimuthal index, which is still meaningful in the present case due to the rotational invariance of the resonator. The label X is a progressive number indicating the position in frequency of a mode with given parity and index $n$. The S modes can be modelled by using a PMC median plane, while N modes require a PEC median plane. The resulting sequence of resonances is indicated in Fig. 6. The $S1_1$ mode is expected at 38.7 GHz, therefore beyond the investigated frequency interval.

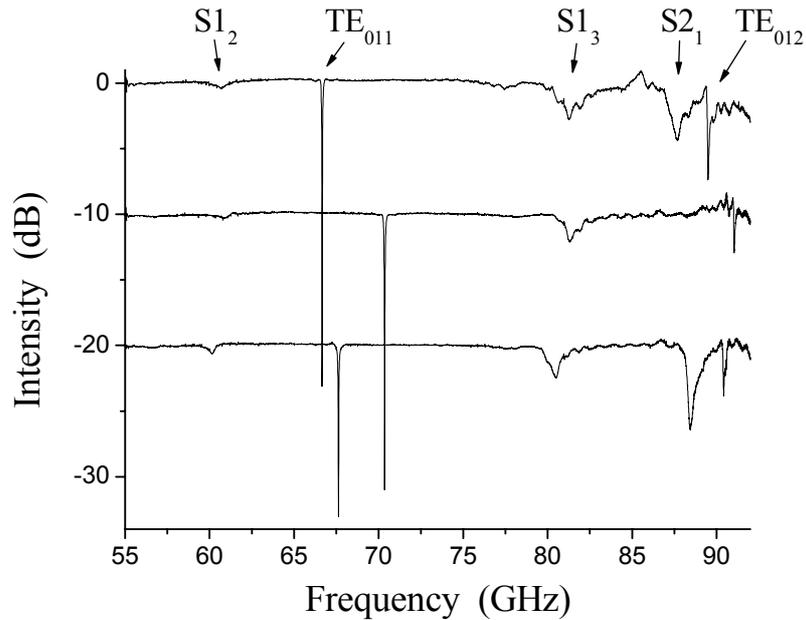

Fig. 6. Normalized spectrum of the resonator shown in Figs. 2 and 3a. Upper curve: homogeneous dielectric disc. Intermediate curve: dielectric disc with a lateral hole. Lower curve: dielectric disc with lateral hole and sample holder. The intermediate curve and the lower one are vertically translated for sake of clarity. The classification of each resonance mode is indicated.

The only resonances with a relevant quality factor are those related to the $TE_0$ modes. However, it has to be noted that the different parameters of the resonator were chosen in order to optimize the $TE_{011}$ mode. The unloaded quality factor of this mode is $Q_0 = 2100$.

The effect of the lateral hole on the spectrum of the resonator can be inferred from the intermediate curve of Fig. 6, which is vertically translated by 10 dB for sake of clarity. The frequency of each resonance was increased by an amount depending on the mode. In the case of the quasi-$TE_{011}$ mode, the perturbed resonance frequency was 70.35 GHz and the related merit factor $Q_0 = 1640$.

The spectrum of the resonator loaded by the sample holder is finally given by the lower curve of Fig. 6, which is vertically translated by 20 dB. The main effect of the capillary, the end of which is aligned with the rim of the disc, is to pull partially back the resonance frequencies. The structure and the position of each resonance are close to those of the initial resonator, so that the field distributions are expected only slightly perturbed. The quasi-$TE_{011}$ mode has a resonance frequency of 67.57 GHz and a merit factor of $Q_0 = 1900$. This frequency can be compared with the expected value as calculated in the previous section, namely 67.26 GHz. The slight degradation of the quality factor observed in the last two configurations can be mainly attributed to the field rearrangement induced by hole and capillary, and by the imperfect planar symmetry.

The influence of the rotations of resonator and sample holder on the resonance frequency of the quasi-$TE_{011}$ mode is reported in Fig. 7. In particular Fig. 7a shows the frequency perturbation due to the variation of $\theta$ in the interval $\left(0, \frac{\pi}{2}\right)$. Almost specular values were obtained in the symmetric interval $\left(\frac{\pi}{2}, \pi\right)$. The frequency perturbation due to the rotation in $\psi$ is reported in Fig. 7b. In both these figures the horizontal lines delimit a region of frequencies which extends over the typical full width at half maximum of the resonance. In almost all cases the observed frequency perturbation was inside the width of the resonance. This perturbation is sensibly more pronounced for rotations in $\psi$, due to the mechanical tolerances that did not ensure a stable position of the capillary along the resonator. A reduction of this perturbation to an order of magnitude similar to that of rotations in $\theta$ is expected for a better realization of the structure. In any case, an automatic frequency control system should easily compensate for these frequency fluctuations.

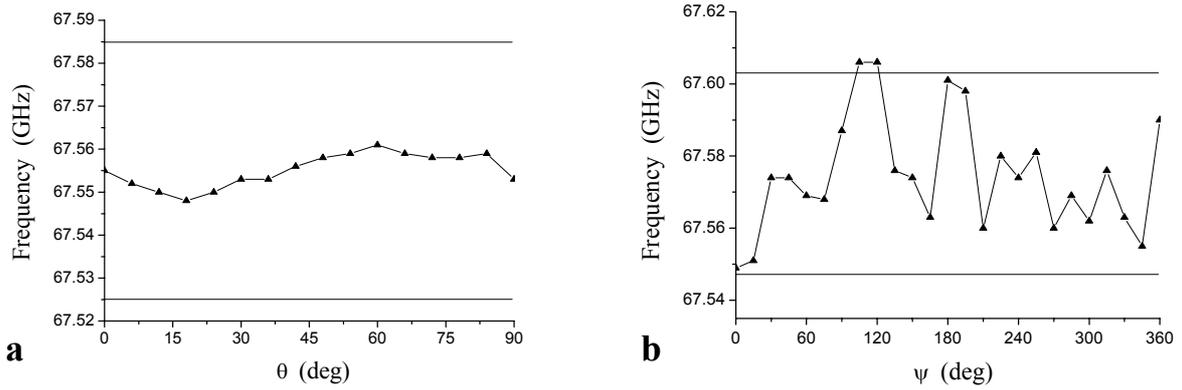

Fig. 7. Resonance frequency perturbations induced by rotations in $\theta$ (**a**) and in $\psi$ (**b**). The horizontal lines mark the half maximum of a fictitious resonance in which the loaded $Q$ is averaged on all measurements.

The other relevant parameters of the resonance were quite stable as well. The unloaded merit factor $Q_0$ oscillated in the range $1860 \pm 150$, which is not far from the measurement accuracy. The coupling level, initially fixed at 13 dB, varied in the interval from 11.3 to 13.5 dB. This variation can be in general easily compensated.

In the second series of measurements the plungers were employed to tune the resonance frequency. Their initial position was fixed at 4 mm from the plane surfaces of the disc. No effects on the resonance frequency and on the quality factor of the quasi-$TE_{011}$ mode were observed, as suggested by the numerical modelling. The plungers were then moved towards the dielectric region. The general behaviour of the tuned resonator was similar to that observed in absence of the plungers. Some differences became appreciable in the conditions of higher tuning. First, some secondary resonances approached to the quasi-$TE_{011}$ mode. Then, the variation of the coupling level as a function of the angle $\theta$ became more and more pronounced, although still compensable by displacing the resonator with respect to the incoming radiation.

## 5. Discussion

The experimental investigation of the previous section confirms the predictions of the theoretical and computational analysis. In particular, a correct design of the proposed resonator leads to a quasi-$TE_{011}$ mode, in which the coupling to the incoming radiation can be easily kept at a constant level during the rotations. This ensures that the intensity of the microwave magnetic field on the sample and its relative orientation with respect to the static magnetic field are independent of the rotation angles.

The crucial point in the design of the resonator is the determination of the field distortion induced by the hole and the sample holder. As shown by the numerical modelling and confirmed by the measurements, when the thickness of the dielectric disc becomes comparable to that of the capillary the perturbed $TE_{011}$ mode loses completely its original rotational invariance, becoming a mode with the electric field concentrated in two lobes external to the capillary. This mode can be employed in principle in EPR applications, since the electric field on the center of the resonator is still quite weak and the magnetic field quite intense. However, in this case the compensation of the coupling level during the rotation in $\theta$ becomes quite problematic, as verified experimentally. In general, the perturbation due to the sample holder can be reduced by using materials with a lower dielectric permittivity, as in case of Teflon. Similar distortions of the field distribution of $TE_{011}$ mode are expected for samples with high permittivity, unless they are confined in the center of the resonator. Analogous care must be used for samples having high dielectric losses. In both cases the design of the resonant structure plays a crucial role in order to minimize undesirable effects.

The experimental results indicate that the performances of a NR resonator laterally loaded by the sample holder are not substantially lower than those of the original, unperturbed resonator. The dependence of the Q factor on the imperfect planar symmetry seems moreover quite weak, allowing thus comfortable mechanical tolerances. These considerations suggest a possible application of the proposed resonators to frequencies sensibly higher than those here investigated. In this case a further simplification of the resonant structure becomes appropriate, in order to avoid the need of too tiny dielectric components. Two possible versions of cylindrical single-mode resonators, based on the same concepts discussed above, are sketched in Fig. 8. In particular, Fig. 8a shows an axially open structure in which the dielectric region is much longer than the employed wavelength. A correct design of this structure leads to a single-mode resonator, as shown in Ref. [15]. The benefit of this configuration is its extreme ease of realization. In particular, the angle $\theta$ can be varied acting directly on the dielectric rod. The tuning of the resonance frequency requires however the displacement of the two conducting plates.

A different kind of open resonator is shown in Fig. 8b. In this case, the starting structure is the single-mode NR cavity discussed in Ref [14]. The benefit of this resonator is the absence of *ad hoc* dielectric regions.

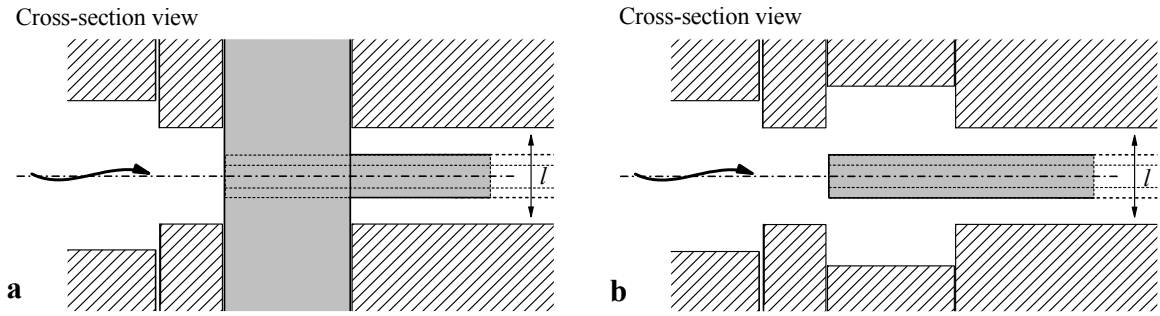

Fig. 8. Cross-section view of different cylindrical NR resonators laterally loaded by the sample holder. **a** Axially open dielectric resonator. **b** Metallic cavity.

Despite the fact that these regions guarantee in general a higher energy density and a more stable mechanical arrangement, the mechanical complexity which arises at very high frequency may prevent any practical benefit. Moreover, the absence of the dielectric regions leads to a larger size of the resonator and allows a direct optical access to the sample. In particular, a cavity with total height of 1.2 mm, diameter of 1.2 mm, and lateral aperture of 0.4 mm resonates at 300 GHz on the $TE_{011}$ mode. When a Suprasil capillary with outer diameter of 0.25 mm and inner diameter of 0.15 mm is inserted in the cavity as shown in Fig. 8b, the resonance frequency decreases to 293 GHz. In this configuration an effective coupling to the incoming radiation is expected for a distance of about 0.2 mm between the rim of the cavity and that of the metallic plates. The resonant structure is thus relatively easy to realize and shows again a quasi-$TE_{011}$ mode only slightly perturbed by the capillary, according to the numerical modelling. On the other hand, the mechanical support of the capillary requires a dedicated structure.

Both the configurations shown in Fig. 8 were investigated from the numerical and the experimental point of view, in the same frequency range of Sect. 4. The results of these investigations are similar to those obtained for the resonator of Fig. 2.

The realization of an EPR spectrometer working at very high frequency requires an analogous simplification of the excitation configuration. A straight geometry, in which the radiation propagates parallel to $\vec{B}_0$, appears as the most appropriate. In this case the sample holder can intercept the propagation structure, so that the excursion of the angle $\theta$ is limited. However, the flange in touch with the resonator can be modified in order to leave just a small 'blind spot' of values around $\theta=0$, where moreover the rotations of the sample become less and less effective.

## 6. Conclusions

To our knowledge, the results here presented demonstrate for the first time that single-mode resonators enabling rotations of the sample about two orthogonal axes can be easily realized also at millimeter wavelengths. The proposed devices are given by cylindrical NR structures in which the sample holder is laterally inserted in the resonant region. The prediction that open configurations without rotational invariance can exhibit confined normal modes follows from symmetry considerations. A proper design of these resonators combines a suitable field distribution to a realistic working configuration. In particular, the relative direction between the microwave field on the sample and the static magnetic field can be independent of the angles of rotation. The compensation of the coupling level during the rotations, which has been proven quite easy in the investigated resonators, ensures a similar independence also of the microwave field intensity. A complete orientation study of the sample is therefore possible at the highest sensitivity and the maximum time resolution, without the need of a dedicated magnet or of remounting of the sample. Although further technical work is still necessary in view of practical applications, the proposed resonators may suggest a decisive innovation in continuous wave and pulsed EPR spectroscopy of anisotropic samples.